Larisa V. Kamaeva[a], Elena N. Tsiok[b,a], Nikolay M. Chtchelkachev[b]

[a] Ural Branch of the Russian Academy of Sciences, Udmurt Federal Research Centre,

Izhevsk. Russia

[b] Russian Academy of Sciences, Vereshchagin Institute for High Pressure Physics, Troitsk,

Moscow, Russia

**Concentration behavior of liquidus temperatures and undercooling of Al-Cu-Co at normal pressure**

Differential thermal analysis has been conducted for the Al-Cu-Co alloys with the composition range of 15 at.% Co and 10 to 30 at.% Cu, and 25 at.% Co and 2.5 to 20 at.% Co. The features of the formation of solid phases have been studied during the crystallization in a crucible in the conditions of slow cooling (rate of cooling to 1 K s$^{-1}$) at normal pressure. On the state diagram of the Al–Cu–Co system with 15 at.% Co and 25 at.% Cu the concentration sections have been built, which allows to determine the concentration ranges from which different phases are formed during the first stage of crystallization. Along the boundaries of different phase regions, extrema are observed on the liquidus line. The observed extrema on the liquidus lines and concentration dependences of undercooling are associated with change in the chemical short-range order at the considered concentrations both in the liquid and solid states.

## 1. Introduction

The Al–Cu–Co system has a stable quasi-crystalline phase (D) with a decagonal structure. The existence of such phases in the solid state is frequently associated with icosahedral short-range order characteristic of metal melts. The possibility of the inheritance of the structural motifs of an initial melt by a solid phase during crystallization and amorphization has been actively discussed in the literature [1, 2]. Our investigations of the Al-Cu-Fe alloys, in which a stable icosahedral quasi-crystalline phase is formed, show that the structure of the melt before the solidification even in the conditions of cooling with the rates to 1 K s$^{-1}$ at normal pressure influences the process of the phase formation especially at the first stages of crystallization [3, 4]. The concentration dependences of the structure-sensitive properties and parameters of chemical short-range order also correlate with the concentration-dependent behavior of the undercooling value and reflect the features of the liquidus lines [3, 5].

As a rule, decagonal quasicrystals have smaller formation energy than icosahedral quasicrystals [6, 7] and they are an intermediate between the quasicrystalline and crystalline states; therefore, the study of the features of the structure inheritance for the systems with decagonal quasicrystals is important for understanding the mechanisms of the phenomenon. The prospects of the practical application of the structure inheritance in the considered system are connected with that in addition to the existence of decagonal quasicrystals, the Al–Cu–Co system is of interest because its Al-rich alloys are very promising structural materials with good casting properties that are provided by the eutectic equilibrium at 540 ºC and 10 at.% Cu [8]. The mechanical properties of above alloys are determined by the morphology and composition of the intermetallic compounds which are strengthening phases [9]. Therefore, the control over the phase-formation processes permits to create alloys with an optimal set of mechanical properties.

In this connection, the goal of the present paper is experimental investigations of phase equilibriums and the processes of the crystallization of the Al–Cu–Co alloys in the region of the D-phase stoichiometry with a large content of Al. However, for the effective use of the structure inheritance phenomenon, only experimental facts confirming the interconnection of the initial melt structure and solid phases formed from it at cooling are insufficient; it is also necessary to create science-based models describing both the melt structure and the possible mechanisms of the structure inheritance between the liquid and solid states, which will permit to predict crystalline structures formed from the liquid state. Therefore, in the present work the structure of the studied melts and changes in the concentration-dependent chemical short-order in the liquid and overcooled states have been numerically modeled as well.

## 2. Experimental procedure

In the present paper, the Al–Cu–Co alloys with the composition ranges of 15 at.% Co and 10 to 30 at.% Cu and 25 at.% Cu and 2.5 to 20 at.% Co were investigated; the concentration step was 2.5 at.%.

The studies of solidification with small rates of cooling were performed by differential thermal analysis (DTA) on an automated device VTA-983 [10, 11]. For the studied samples, the DTA curves (thermograms) – the dependences of the temperature difference between a reference and a sample on the temperature of a reference – were obtained by heating the samples after remelting to 1400 or 1500 ºC with the rate ($v$) of 20 K min$^{-1}$; after that the samples were held at the mentioned temperatures for 20 min and finally cooled with the rate of 100 K min$^{-1}$. Based on the DTA data the temperatures of all the stages of melting (in the mode of heating) and crystallization (in the mode of cooling) were determined. The value of undercooling ($\Delta T$) was determined by the difference between the liquidus temperature and the initial crystallization temperature. Additionally, two series of the experimental studies on cycling and thermal cycling with different rates were carried out for the thorough analysis of the influence of the temperature and the cooling rate of the melt on the overcooling value. In the experimental study on thermal cycling each cycle consisted of the sample heating to a specified temperature and cooling with the rate of 100 K min$^{-1}$ after holding for 20 min. The measurements during the first cycle were made at heating to 1400 ºC or 1500 ºC depending on the melt composition and subsequent cooling; after that the overheating temperature was lowered by 10–30 ºC and so on to the liquidus temperature. Based on the obtained thermograms, the overcooling values were calculated, and the equilibrium temperatures of the liquidus (at heating) and the initial crystallization temperature in the conditions of cooling from different temperatures were determined. As a result, the dependences of overcooling, in the condition of which the melt crystallization started, on the melt temperature and cooling rate were obtained.

The data on the features of the atomic short-range order in the melts were obtained by molecular-dynamics modeling based on a complex approach; the approach includes quantum molecular-dynamics calculations and machine learning. The ab initio calculations were performed using the Vienna ab initio simulation modeling package (VASP) [12]. For the machine learning, a DeePMD package [13] and a concurrent learning platform DP-GEN [14] were used.

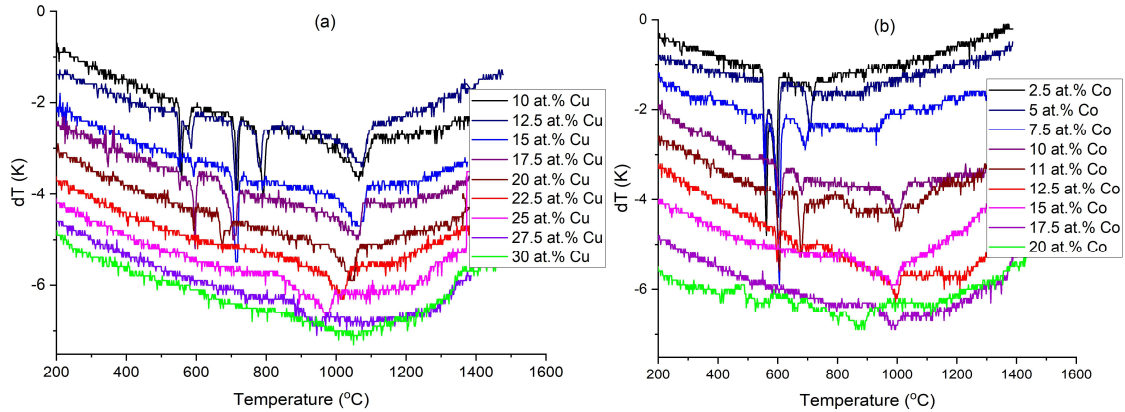

Fig. 1. DTA curves (thermograms) during heating at constant rate ( v = 20  K min-1) of the Al–Cu–Co alloys with 15 at.% Co (a) and 25 at.% Cu (b), the endothermic effect is directed downwards.

### 3. Results and discussion

Figure 1 shows the thermograms of heating of the studied samples. Based on the termograms, the temperatures of the phase transitions taking place in the studied samples, including the liquidus and solidus, and their concentration variations are precisely determined (Fig. 2a, b). The analysis of the thermograms of heating shows that an extended two-phase region is observed for all the samples and the temperature difference between the solidus and liquidus exceeds 500 ºC for some compositions. The number of endothermic effects depends on the Cu and Co concentrations and varies from 2 to 7. The alloys with 15 at.% Co and Cu from 10 to 17.5 at.% have an expressed peak on the thermograms of heating; the value and shape of the peak indicate eutectic melting with a large thermal effect. On the DTA curves several additional small endothermic peaks are also observed corresponding to different peritectic transformations. On the heating thermogram of the alloy with 20 at.% Cu the magnitude of the effect associated with the melting of eutectic decreases and the second stage of melting becomes more expressed. For the alloys with the Cu content from 20 to 30 at.%, the thermograms of heating undergo more essential changes; there are no expressed peaks on the thermograms and only kinks are observed  indicating the temperatures of the solidus and liquidus. At 25 at.% Cu, the peak corresponding to the eutectic melting is observed for the alloys with the Co content from 2.5 to 7.5 at.%. On the heating thermogram of the alloy with 10 at.% Co the magnitude of the effect associated with the eutectic melting decreases and the second stage of melting becomes more expressed.

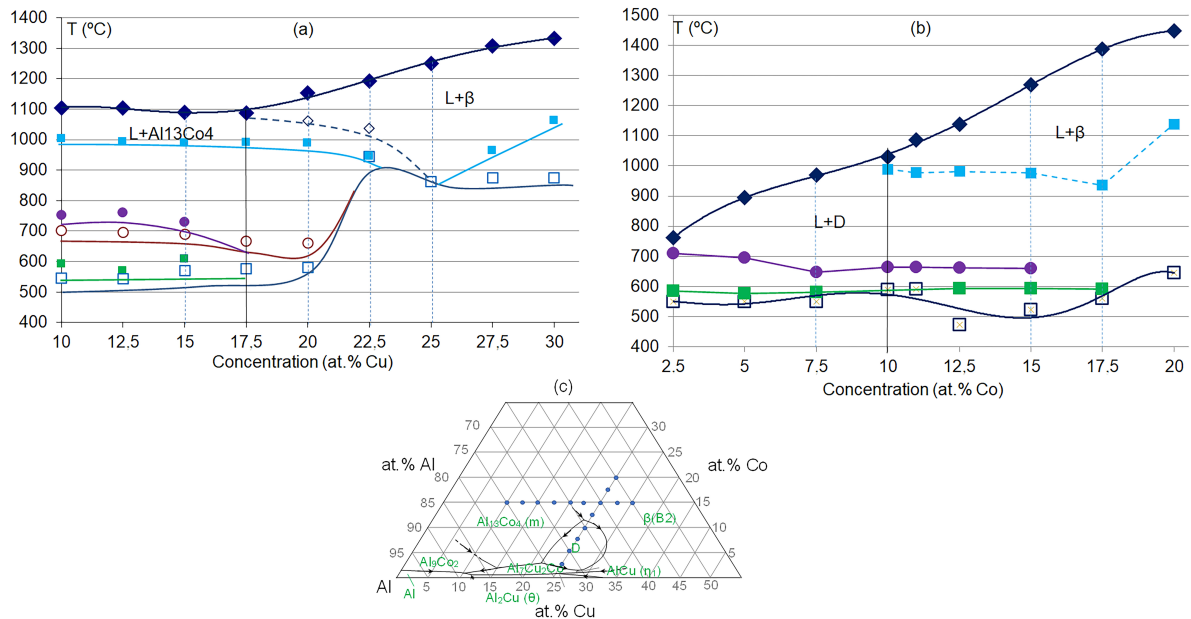

Fig. 2. Concentration dependences of the temperatures of the phase transformations in the Al–Cu–Co system with 15 at.% Co and 10–30 at.% Cu (a); with 25 at.% Cu and 2.5–20 at.% Co (b); a projection of the liquidus surface based on the data from [8] where the studied alloys are shown by points..

The analysis of the DTA data on heating allows building the concentration dependences of the temperatures of the phase transformations in the Al–Cu–Co alloys with 15 at.% Co (10–30 at.% Cu) and 25 at.% Cu (2.5–20 at.% Co) presented in Fig. 2a, b. Additionally, in Fig. 2 a projection of the liquidus surface in the studied concentration region of the state diagram according to the data from [8] is shown. Figure 2 is the basis for building the corresponding concentration sections on the state diagram. The concentration ranges in which the character of the thermograms of the Al–Cu–Co alloys changes and the concentration region with different phase composition below the liquidus line are indicated by dotted lines. The phase composition of the regions is determined by comparing the obtained results and the literature data on the phase equilibriums in the Al–Cu–Co system [8, 9]. Thus, we determined the concentration-dependent behavior of the temperatures of the liquidus at 15 at.% Co (10–30 at.% Cu) and 25 at.% Cu (2.5–20 at.% Co). The built lines of the liquidus have kinks at the concentrations at which the change in the form of the solid phase melting takes place. At the variation of the Al and Cu concentrations and 15 at.% Co a kink is observed at 20 at.% Cu and it corresponds to the invariant transformation (Fig. 2):

Al$_{13}$Co$_4$ + L → $\beta$ + L (1)

For the second concentration section, at 25 at.% Cu in the region of 10 at.% Co the change in the form of the solid-liquid equilibrium occurs. Up to 10 at.% Cu it is a D-phase, and at further increase of the Co content a $\beta$-phase – a solid solution based on the Cu$_3$Al – appears.

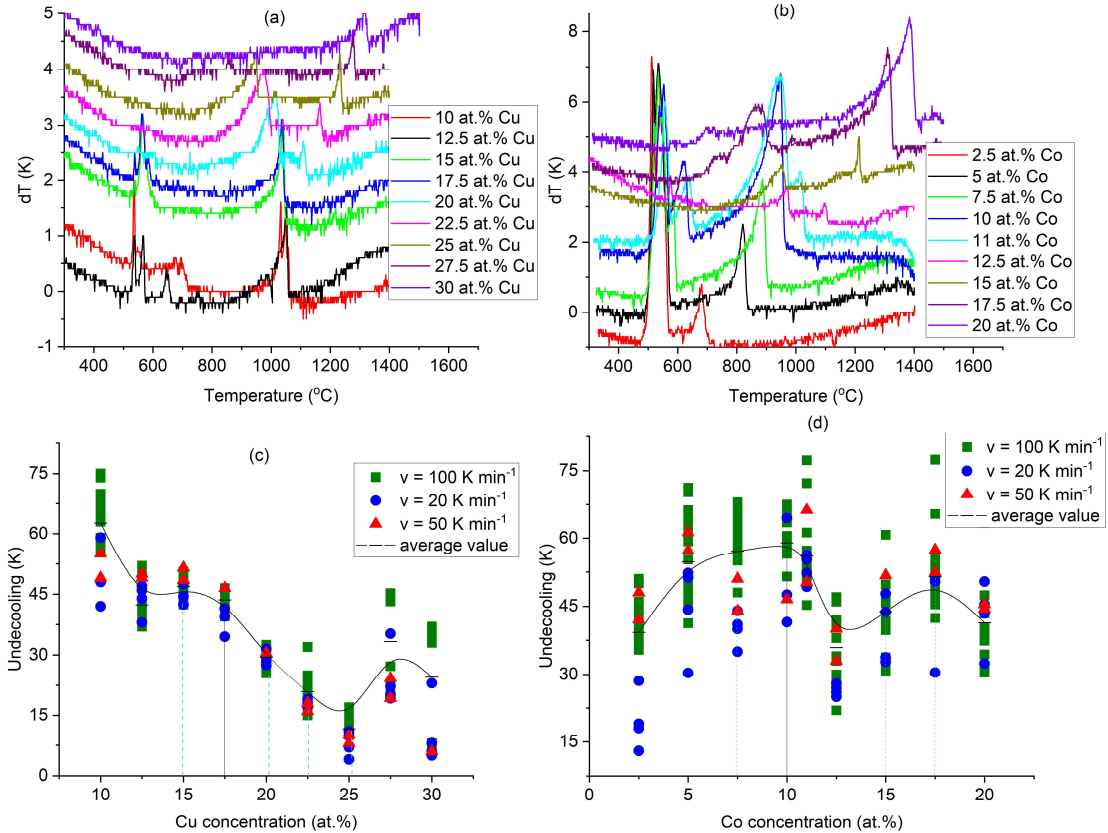

Fig. 3. Thermograms of cooling (v = 100 K min-1) (a, b) and the concentration dependences of undercooling (c, d) of the Al–Cu–Co alloys with 15 at.% Co (a, c) and 25 at.% Cu (b, d).

Figure 3 shows the thermograms of cooling at the cooling rates of 100 K min$^{-1}$ and the concentration dependences of undercooling for the studied samples. It can be seen from the figure that the alloy crystallization starts in the conditions of small undercooling by 5–100 K and depends on the concentrations and cooling rates; the maximum overcooling is observed for the alloys Al$_{75}$Cu$_{10}$Co$_{15}$ and Al$_{64}$Cu$_{25}$Co$_{11}$. Both at cooling and at heating, the thermograms differ for the compositions from different concentration regions of the diagram; therefore, for each of the two concentration sections, three concentration ranges can be distinguished, which have different types of crystallizations: 1) for 15 at.% Co, the ranges are 10–12.5, 15–17.5, and 20–30 at.% Cu; 2) for 25 at.% Cu, the ranges are 2.5–7.5, 10–15, and 15–20 at.% Co. The regions are in good agreement with the DTA results obtained in the mode of heating (Figs. 1, 2), which indicates that in the selected conditions the character of the alloy crystallization is close to that of equilibrium. At the Al-based alloy crystallization in the corundum ($Al_2O_3$) crucible, the undercooling value can be considerably influenced by different factors such as cooling rate, melt temperature, and the number of the cycles 'melting-crystallization' [15]. Therefore, a series of measurements has been carried out for each sample. The measurements consist of several cycles 'heating (melting) - cooling (crystallization)', in which either the maximum temperature of the sample heating before cooling or the rate are varied. The experiments have been conducted at the cooling rates of 20, 50 and 100 K min$^{-1}$. All the obtained data have been used for determining the undercooling value at which the studied alloy crystallization starts. The summarized concentration dependences are shown in Fig. 3c, d. The studies show that the cooling rate influences the undercooling value, at which the crystallization starts, and the sequence and character of the crystallization stages. At the increase of the cooling rate, the initial crystallization temperature decreases and the undercooling value increases; such influence is maximally manifested for the Al-rich melts with a small content of Co (Fig. 3d). The change of the thermogram shapes at the cooling rate variation is observed in the low-temperature region near the solidus temperature. At cooling with the rate of 100 K min$^{-1}$, the melt temperature does not influence the undercooling value for most of the studied alloys. The exceptions are the alloys having more than 12.5 at.% Co at 25 at.% Cu. For such alloys the increase of the melt temperature leads to the growth of overcooling at which the β-phase crystallization starts. However, despite the considerable dispersion of the values the tendency of the concentration influence on overcooling is observed (Figs. 3c, d).

The concentrations, at which the change of the type of crystallization is observed, can be seen on the curves of the concentration dependences of undercooling; however, the absolute value of undercooling is determined, first of all, by the interaction in the melt – crucible – atmosphere system. The obtained data show that the conditions of equilibrium for the studied system depend on the component content in the melt.

The curves of the concentration dependences of overcooling have a more complex shape than the lines of the liquidus. The features revealed on the curves of the overcooling concentration dependence are in good agreement with the general change of the character of crystallization (with the shapes of the thermograms of cooling). Taking into consideration that the sequence of the phase formation during solidification corresponds to the state diagram, the obtained concentration dependences of cooling can be divided into two main ranges: in the selected conditions of cooling, for the composition range of 15 at.% Co and 10 to 20 at.% Cu the concentration dependence of overcooling describes the changes in the crystallization ability for the $Al_{13}Co_4$ intermellide and for the composition range of 15 at.% Co and 20 to 30 at.% Cu – the changes in the crystallization ability for the β-phase (Fig. 3c); for the composition range of 25 at.% Cu and 2.5 to 10 at.% Co, the concentration dependence of overcooling describes the changes of the crystallization ability for the D-phase, and from 10 to 20 at.% Co – for the *β*-phase (Fig. 3d). To analyze the influence of the chemical interaction in the overcooled melt on the concentration dependences ΔT, the investigation of the chemical short-range in the Al-Cu-Co melts has been performed at 727 ºC.

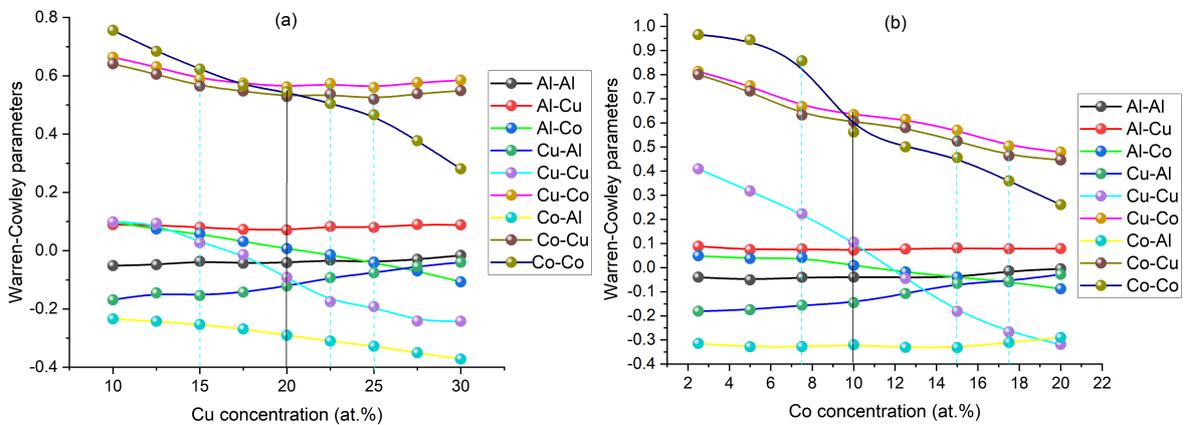

Fig. 4. Concentration dependences of the Warren-Cowely parameters for the overcooled (T = 727 ºC) Al–Cu–Co melts with 15 at.% Co (a) and 25 at.% Cu (b).

As a result of the complex molecular dynamics modeling with the use of ab initio molecular dynamics and machine learning, the full and pair correlation functions of the radial distribution (FRDs) of atoms have been calculated for the studied alloys. The main characteristics of the short-range order, including the Warren-Cowely parameter ($\alpha$), have been determined with the use of the full and pair FRDs of atoms [4, 5, 11]. The $\alpha$-parameter shows how much difference there is between the chemical surroundings of the selected atom i and the random distribution of atoms of the j-type. If $\alpha$(i-j) = 0, the number of the j-atoms surrounding the i-atom is determined by the concentration of the j-atoms in the melt with regard for the general number of atoms in the first coordination shell. The negative $\alpha$(i-j) indicates an effective attraction between the atoms of the i and j types, i.e. the number of the atoms of the j-type in the surroundings of i atom is larger than that at the random distribution; the positive $\alpha$(i-j) corresponds to an effective repulsion. The obtained results are given in Fig.

4. As seen from Fig. 4, the largest chemical interaction influencing the structure in the overcooled Al–Cu–Co melts is observed between the Co and Cu atoms in the region of the large Al concentrations. Around the Co and Cu atoms, the effective repulsion between the pairs Co-Co, Co-Cu, Cu-Co and Cu-Cu is characterized by not only large values of $\alpha$ but also their expressed concentration dependence, namely, the higher the Co or Cu concentration is in the melt, the lower the values of the Warren-Cowely parameters are. Such tendency leads to that the parameters $\alpha$(Cu-Cu) change the sign at 15 at.% Co (Fig. 4a) and 12.5 at.% Cu (Fig. 4b). The observed character of the concentration-dependent change of the interaction between the atoms of the transition metals in the overcooled melts is in good agreement with the concentration dependence of the liquidus temperatures, which indicates the conservation of such interaction in the solid phase as well. The effective repulsion between the Co and Cu atoms leads to that the formation of ternary intermetallic phase is not observed at high temperatures in the Al–Cu–Co system. For the studied melts, the main tendencies of the concentration-dependent behavior of overcooling are determined by the character of the change of the interaction between the Al and Co atoms observed around the Co atoms; from this it can be concluded that in the Al–Cu–Co melts, during crystallization clusterization starts around the Co atoms. The obtained results are in good agreement with the earlier obtained data for the Al–Cu–Fe and Al–Cu–Ni systems [4, 11].

**Conclusion**

Thus, based on the results of the differential thermal analysis of the Al–Cu–Co alloys with the composition ranges of 15 at.% Co and 10 to 30 at.% Cu and 25 at.% Cu and 2.5 to 20 at.% Co, the concentration sections have been built on the state diagram for the Al-Cu-Co system with 15 at.% Co and 25 at.% Cu. This permits to determine the concentration ranges from which different phases are formed at the first stage of crystallization. Along the boundaries of different phase regions, extrema on the line of the liquidus are observed. The studies of the crystallization processes for the Al–Cu–Co alloys in the conditions of slow cooling with the rates down to 1 K s$^{-1}$ in the Al$_2$O$_3$ crucibles show that the alloy crystallization starts in the conditions of small undercooling by ~50 K and depends on the conditions of cooling and the concentration of Cu and Co. For the alloys with 25 at.% Cu and the Co content of no less than 10 at.% the maximum influence on the overcooling value is produced by the cooling rate. The concentration dependences of overcooling are non-monotonic and the observed extrema do not correlate with the features of the lines of the liquidus. The comparison of the obtained experimental data with the results of the complex molecular dynamics modeling with the use of ab initio molecular dynamics and machine learning shows that the largest chemical interaction influencing the structure in the overcooled Al–Cu–Co melts is the effective repulsion around the Co and Cu atoms between the Co-Co, Co-Cu and Cu-Co pairs; changes in their concentration determine the character of the liquidus lines. For the studied melts, the main tendencies of the concentration-dependent behavior of overcooling are determined by the character of the change of the effective attraction between the Al and Co atoms.

The investigation has been financed by the grant of the Russian Science Foundation No. 22-22-00912, https://rscf.ru/project/22-22-00912/